\documentclass[usegraphicx,usenatbib]{mn2e}

\newcommand{\cpetro}{{\ensuremath{C_{\mathrm{Petro}}}}}
\newcommand{\cpetrodisc}{{\ensuremath{C_{\mathrm{Petro}}^{\mathrm{disc}}}}}
\newcommand{\cnorm}{{\ensuremath{C_{\mathrm{norm}}}}}
\newcommand{\sech}{{\ensuremath{\mathrm{sech}}}}
\newcommand{\kms}{{\ensuremath{\mathrm{km~s^{-1}}}}}
\newcommand{\absmag}{{\ensuremath{{}^{0.1}M_r}}}
\newcommand{\CMD}{{\ensuremath{\mathrm{CMD}}}}
\newcommand{\sersic}{S\'{e}rsic}

\title[Galaxy Concentrations are Trimodal]{Galaxy Concentrations are Trimodal}
\author[Bailin and Harris]{Jeremy Bailin\thanks{bailinj@mcmaster.ca},
 William E. Harris \\
Department of Physics and Astronomy, McMaster University, 1280 Main Street West,
Hamilton, ON, L8S 4M1, Canada}

\begin{document}

\maketitle

\begin{abstract}
We have analysed the distribution of inclination-corrected
galaxy concentrations in the Sloan Digital Sky Survey. We find that unlike most
galaxy properties, which are distributed bimodally, the
distribution of concentrations is trimodal: it exhibits three distinct peaks.
The newly-discovered intermediate peak, which consists of
early-type spirals and lenticulars, may contain $\sim 60\%$ of the number
density and $\sim 50\%$ of the luminosity density of $\absmag < -17$
galaxies in the local universe.
These galaxies are generally red and quiescent, although 
the distribution contains a tail of blue star-forming galaxies
and also shows evidence of dust.
The intermediate-type galaxies
have higher apparent ellipticities than either disc or elliptical
galaxies, most likely because some of the face-on intermediate types are
misidentified as ellipticals.
Their physical half-light radii are
smaller than the radii of either the disc or elliptical galaxies, which may be
evidence that they form from disc fading.
The existence of a distinct peak in parameter space associated with
early-type spiral galaxies and lenticulars implies that
they have a distinct formation mechanism and are not simply the
smooth transition between disc-dominated and spheroid-dominated galaxies.
\end{abstract}


\section{Introduction}
Morphological classification of galaxies dates back to the \citet{hubble26}
``Tuning Fork'' that classified galaxies into ellipticals and spirals
(and further into barred and unbarred spirals). More recent and quantitative
classifications of large numbers of galaxies from current galaxy redshift surveys
have upheld this bimodality, finding that morphological
properties such as the global concentration of the light
(e.g. as measured by the Petrosian concentration parameter \cpetro),
the functional
form of the surface brightness profile (e.g. exponential versus
$R^{1/4}$ law, or the value of the \citet{sersic-profile} $n$ index),
and the asymmetry are best decomposed into two populations: concentrated,
smooth, de~Vaucouleurs (or high $n$) ellipticals and
diffuse, clumpy, exponential spirals.
Properties
of galaxies sensitive to their star formation histories,
such as the global colour, the presence of nebular emission lines,
the presence
of spectral features symptomatic of young stars, and the detailed
stellar populations when they can be resolved, are also decomposed
into two populations:
red, quiescent ellipticals full of old stars and blue, star-forming
spirals with young stellar populations
\citep[see][and references therein]{shimasaku-etal01,strateva-etal01,
blanton-etal03-properties,baldry-etal04,driver-etal06}.

At the joining point of the tuning fork lies the S0 or lenticular class.
These are galaxies that are too flattened not to
have a disc morphology, but that
appear dominated by their spheroidal component and appear more similar
to ellipticals than discs in their colour and environment.
These intermediate galaxies may be a unique class of
object with a particular formation scenario,
such as disc fading,
or they may simply represent the
smooth transition from galaxies with low bulge-to-disc
ratios, $B/D$, to those with high $B/D$.

If lenticulars have a unique formation scenario, they are likely to
congregate in a particular region of galactic parameter space.
In this paper, we test this possibility by studying the morphological
distribution of galaxies in the Sloan Digital Sky Survey.
Section~\ref{sdss-section} details the galaxy sample.
In Section~\ref{trimodality}, we investigate the distribution of
galaxy concentrations and find that they are trimodal.
In Section~\ref{intermediate properties}, we investigate the
properties of the newly-discovered intermediate class of galaxies.
In Section~\ref{discussion} we discuss the implications
of our results, and Section~\ref{conclusions} contains
our conclusions.

\section{Selection of SDSS Galaxies}
\label{sdss-section}
The Sloan Digital Sky Survey (SDSS) is a 5-band optical and near-infrared
imaging and spectroscopic survey covering one quarter of the sky.
\citep{sdss-technical-summary}.
We use data from Data Release 6 \citep[DR6;][]{sdss-dr6}.
Our sample consists of the $543010$ galaxies that meet the
Main Galaxy Sample targeting criteria \citep{sdss-maingalaxysample},
and have spectra that are classified as galaxies with
confident redshifts ($z_{\mathrm{conf}} > 0.9$).
Petrosian magnitudes are used throughout,
and $r$-band quantities are used for all photometric parameters.
K-corrections and $1/V_{\mathrm{max}}$ terms are calculated
using KCORRECT v4\_1\_4 \citep{kcorrect}.
We assume $\Omega_0=0.3$, $\Omega_{\Lambda}=0.7$ and
$H_0=70~\kms~\mathrm{Mpc^{-1}}$ when calculating distance moduli and
angular diameter distances.

\section{Morphological Trimodality}
\label{trimodality}
Intrinsic galaxy properties have been shown to be distributed bimodally.
Measurements that are sensitive to the star formation properties of
galaxies indicate that they are either (1) red, not currently
forming stars, and have passively-evolving stellar populations,
or (2) blue and actively star forming, while relatively
few galaxies have intermediate properties.
Similarly, morphological measurements, such as concentration,
surface brightness profile, and visual appearance indicate that galaxies
are either concentrated smooth elliptical galaxies with de~Vaucouleurs-type
profiles, or that they are extended disc galaxies with spiral structure
and exponential profiles.

Here we focus on the $r$-band Petrosian concentration \cpetro, defined
to be the ratio
\begin{equation}
\cpetro \equiv \frac{R_{90}}{R_{50}},
\end{equation}
where $R_{50}$ and $R_{90}$ are the radii enclosing $50\%$ and
$90\%$ of the Petrosian flux respectively.
Examination by eye of samples of galaxies with large
values of \cpetro, usually considered to be early-types, reveals
that many appear to be disc systems.
One way to further distinguish elliptical
systems from discs is to consider their apparent axis ratio. Ellipticals,
which are approximately spherical, cannot
appear to have small apparent axis ratios from any viewing angle.
Disc galaxies, on the other hand, can appear very flattened when
viewed edge-on.

\begin{figure}
\includegraphics[scale=0.5]{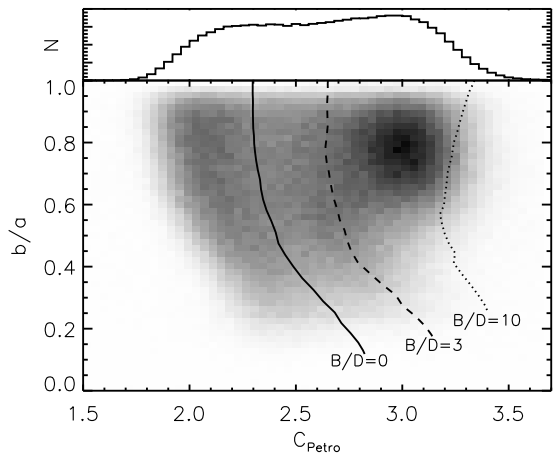}
\caption{\label{cpetro-vs-ba}%
(Top) Histogram of the Petrosian concentration, \cpetro, for SDSS galaxies.
(Bottom)
Distribution of SDSS galaxies in the \cpetro\ vs. isophotal
axis ratio, $b/a$, plane.
Solid, dashed and dotted lines indicate the loci
of dust-free galaxies with bulge-to-disc ratios, $B/D$, of $0$,
$3$ and $10$ respectively
(see Appendix~\ref{model-bd-systems} for details).}
\end{figure}

In Figure~\ref{cpetro-vs-ba}, we plot the distribution of galaxies in the
$\cpetro$-$b/a$ plane, where $b/a$ is the ratio of the isophotal minor
and major axes. Three features are apparent in this diagram:
\begin{itemize}
 \item At the left, there is a band of galaxies extending from 
$\cpetro \approx 2$ and $b/a \approx 0.9$ to $\cpetro \approx 2.5$
and $b/a \approx 0.2$. These low-concentration galaxies that extend to
very extreme axis ratios are the classical late-type disc galaxies.
 \item At the top right, there is a cloud of galaxies centered at
$\cpetro \approx 3$ and $b/a \approx 0.8$. These high-concentration
round galaxies are the classical early-type elliptical galaxies.
 \item Intermediate between these classes,
there is a strip of galaxies extending
from $\cpetro \approx 2.6$, $b/a \approx 0.65$ to
$\cpetro \approx 3$, $b/a \approx 0.2$, approximately along
the $B/D=3$ curve. It is separated from both
the late-type strip and early-type cloud by gaps, regions of parameter
space that are much less populated.
\end{itemize}

We have also plotted on Figure~\ref{cpetro-vs-ba}
a histogram of the concentration
parameters. Although the distribution of \cpetro\ is
bimodal, the ``early-type'' peak at high \cpetro\ contains
galaxies belonging
both to the early-type cloud and to the intermediate-type strip.

The late-type and intermediate-type strips, both of which extend
over a large range in axis ratio, follow a tilted locus in
this plane: galaxies with smaller axis ratios are more concentrated.
A similar trend among late-type disc galaxies (Sb and later) was
noted and quantified by
\citet{yamauchi-etal05}, who studied a sample of $1817$ galaxies
drawn from SDSS DR1 \citep{sdss-dr1}.
Because there should be
no difference between the intrinsic properties of galaxies with
different apparent axis ratios\footnote{At least for disc galaxies, which
are more or less axisymmetric and so differences in apparent axis ratio
simply reflect different inclinations of the disc},
this trend must be purely an inclination effect;
indeed, when \citet{yamauchi-etal05}
calculated the Petrosian concentration using elliptical apertures instead
of circular apertures, they found that it was independent
of axis ratio.
Rather than use their method, which would require re-analysing the
images of each spectroscopic galaxy in DR6, we model the effect
of inclination by constructing models of bulge+disc systems viewed
at a variety of inclination angles and determining their apparent
axis ratios, $b/a^{\mathrm{disc}}$, and concentrations, \cpetrodisc.
(see Appendix~\ref{model-bd-systems} for details).
The relation we find for a pure exponential disc with no bulge is plotted
as the solid curve labelled $B/D=0$ in Figure~\ref{cpetro-vs-ba}.
The shape of this line matches 
the shapes of both the late-type and intermediate
strips in the $\cpetro$-$b/a$ plane.
The disc locus is offset to slightly lower values of \cpetro; the
reason for this is investigated in Appendix~\ref{concentration-section}.
Using the relation between $b/a^{\mathrm{disc}}$ and \cpetrodisc,
we define for each galaxy $i$ with observed concentration $\cpetro_i$
and axis ratio $(b/a)_i$,
a ``normalised'' concentration, $\cnorm$:
\begin{equation}
  \cnorm_i \equiv \frac{\cpetro_i}{\cpetrodisc(b/a_i)}.
\end{equation}

\begin{figure}
\includegraphics[scale=0.5]{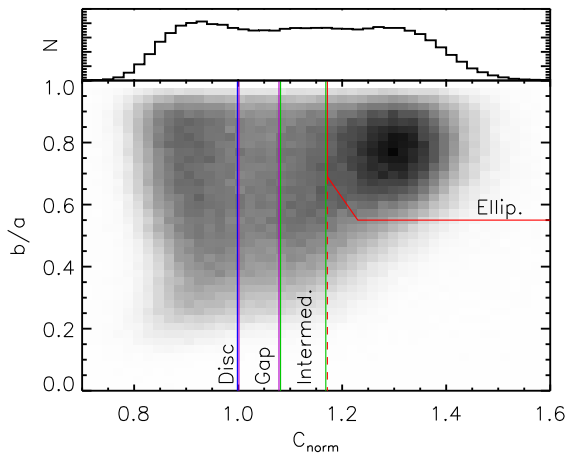}
\caption{\label{cnorm-vs-ba}%
As in Figure~\ref{cpetro-vs-ba} for the normalized concentration \cnorm.
The coloured lines delineate the ``Disc'' (blue), ``Gap'' (purple),
``Intermediate'' (green) and ``Elliptical'' (red) regions of the
\cnorm-$b/a$ plane. When we require samples unbiased in $b/a$, we
include all galaxies with $\cnorm > 1.17$ in the Elliptical sample,
as indicated by the red dashed line.}
\end{figure}

The distribution of galaxies in the $\cnorm$-$b/a$ plane is shown in
Figure~\ref{cnorm-vs-ba}, along with a histogram of \cnorm\ values.
The strong variation of \cpetro\ with axis ratio is
almost entirely eliminated when we
use \cnorm, validating its use as an inclination-independent measure
of galaxy concentration.

The distribution of \cnorm\ indices is broad and appears at first
glance to be multimodal.  
To carry out quantitative fits to the concentration
distribution, we employed the
RMIX package \citep{macd07}, a statistics code written with the 
programming language R.  RMIX is designed to find best-fit combinations
of multimodal histograms that can be built up from combinations 
of simple functional components (the allowed model functions include
Gaussian, Poisson, gamma distributions among others).
The number of components adopted
to fit the data is user-defined, as are the 
means and variances of the model components, any of which can be constrained
or determined by the fit, in arbitrary combinations as appropriate.
\footnote{The code is available from Peter MacDonald's website at
http://www.math.mcmaster.ca/peter/mix/mix.html.}

\begin{figure}
\includegraphics[scale=0.44]{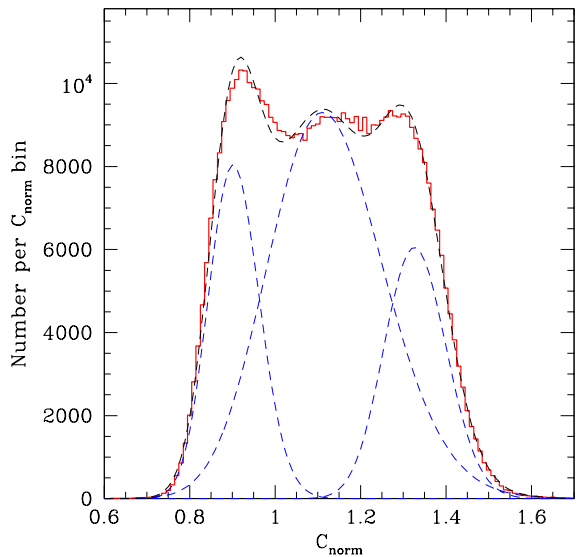}
\caption{\label{trimodal_fit}%
Trimodal fit to the observed \cnorm\ distribution, shown as the red solid
histogram with bins of width $0.01$.
The upper black dashed line indicates the
best fit sum of three gamma distributions, which has a goodness-of-fit
equal to $\chi_{\nu}^2 = 0.77$, while the lower blue dashed lines indicate
the individual components in the fit.}
\end{figure}

In our case, the most basic question we are interested in is this:
{\sl what is the minimum number of simple components needed to describe
the \cnorm\ distribution?}  It can quickly be found that no {\sl bimodal}
model built from any of these simple unimodal functions can fit the
distribution adequately (typical goodness-of-fit values for
bimodal models are 
$\chi_{\nu}^2 \sim 10$).
A trimodal Gaussian model performs much better at 
$\chi_{\nu}^2 = 1.73$.  A four-mode Gaussian does still better, reaching
the optimum $\chi_{\nu}^2 = 1.0$.  
However, we have no particular reason aside from
numerical convenience to adopt the Gaussian model.
Another almost equally convenient model is the standard gamma distribution,
given by
\begin{equation}
g(x) \, = \, x^{\alpha-1} { {\beta^{\alpha} e^{-\beta x}} \over {\Gamma(\alpha)}},
\end{equation}
where ($\alpha,\beta$) are parameters determining the shape and degree of
symmetry.
The function has a mean $\mu = \alpha / \beta$
and variance $\sigma^2 = \alpha / \beta^2$.
It is a more versatile function allowing for modest asymmetry.
This extra degree of freedom allows us to match the data with a trimodal
model and an optimum goodness-of-fit equal to 
$\chi_{\nu}^2(gamma) = 0.77 $, superior to the 4-mode Gaussian%
\footnote{We note in passing that a lognormal and trimodal distribution
does almost as well. From a numerical standpoint, the main conclusion
is that models built from standard fiducial functions that
are unimodal but allow for some asymmetry can provide excellent
fits to the data.}
This result is shown in
Figure \ref{trimodal_fit}.  The mean \cnorm\ values for the three modes
are (0.906, 1.127, 1.330), standard deviations (0.059, 0.135, 0.073)
and proportions (0.220, 0.577, 0.203).  The sample size is so large that
each parameter is internally uncertain to less than 1\%.
The parameters of these fits are given in Table~\ref{rmix fit table}

\begin{table}
\caption{Parameters of Trimodal Fits\label{rmix fit table}}
\begin{tabular}{lccc}
\hline
{Peak} & {Fraction} & {Mean} & {Standard Deviation}\\
\hline
\multicolumn{4}{c}{Raw Number Counts}\\
Low & $0.220\pm0.007$ & $0.906\pm0.001$ & $0.059\pm0.001$\\
Intermediate & $0.577\pm0.007$ & $1.127\pm0.004$ & $0.135\pm0.001$\\
High & $0.203\pm0.007$ & $1.330\pm0.002$ & $0.073\pm0.002$\\
\\
\multicolumn{4}{c}{Comoving Number Density}\\
Low & $0.366\pm0.019$ & $0.901\pm0.002$ & $0.061\pm0.001$\\
Intermediate & $0.581\pm0.025$ & $1.076\pm0.005$ & $0.114\pm0.004$\\
High & $0.053\pm0.012$ & $1.311\pm0.013$ & $0.085\pm0.005$\\
\\
\multicolumn{4}{c}{Comoving Luminosity Density}\\
Low & $0.287\pm0.004$ & $0.907\pm0.001$ & $0.061\pm0.001$\\
Intermediate & $0.519\pm0.007$ & $1.115\pm0.002$ & $0.118\pm0.001$\\
High & $0.194\pm0.005$ & $1.331\pm0.001$ & $0.078\pm0.001$\\
\\
\multicolumn{4}{c}{$R_{50} > 1.75\arcsec$ Subsample}\\
Low & $0.292\pm0.007$ & $0.909\pm0.001$ & $0.061\pm0.001$\\
Intermediate & $0.498\pm0.006$ & $1.103\pm0.004$ & $0.140\pm0.002$\\
High & $0.210\pm0.007$ & $1.344\pm0.001$ & $0.072\pm0.002$\\
\end{tabular}
\end{table}

The strong conclusion from this analysis is that, if we describe the 
\cnorm\ distribution by assuming that it is built up from a combination
of unimodal components with relatively simple forms, then
bimodal solutions are strongly ruled out and
{\sl three} components are
are a necessary and sufficient condition. This is in contrast to the
bimodal distribution usually associated with many other galaxy properties.
We identify the galaxies in the middle peak
(the green strip in Figure~\ref{cnorm-vs-ba})
as ``intermediate'' type galaxies that are structurally
distinct both from early- and late-type galaxies.

\begin{figure}
\includegraphics[scale=0.5]{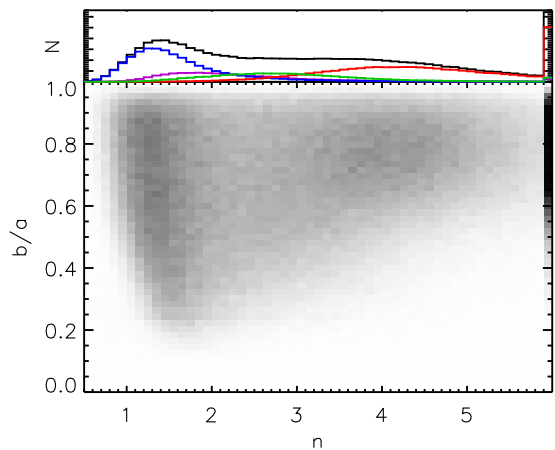}
\caption{\label{sersic figure}%
As in Figure~\ref{cpetro-vs-ba} for the \sersic\ $n$ index.
The coloured histograms show the marginalized distributions of $n$
for the different populations
identified in Figure~\ref{cnorm-vs-ba}: disk (blue), gap (purple),
intermediate (green) and elliptical (red).
The peak at $n\approx 6$ is an artefact of the NYU-VAGC profile fitting
procedure.}
\end{figure}

Because the Petrosian concentration parameter sometimes behaves
unintuitively (see Appendix~\ref{concentration-section}), it is
worth confirming that other measures of galaxy concentration
also exhibit trimodality. One common measure is the
\citet{sersic-profile} $n$ index, which equals $1$ for a pure exponential
profile and $4$ for a \citet{devaucouleurs48} profile. The
NYU Value Added Galaxy Catalog \citep[NYU-VAGC][]{nyuvagc-dr2}
contains \sersic\ profile fits to most galaxies in the SDSS. We have
cross-matched the version of NYU-VAGC based on SDSS DR6%
\footnote{http://sdss.physics.nyu.edu/vagc/} against
the galaxies in our sample. The distribution of galaxies
in the $n$-$b/a$ plane is shown in Figure~\ref{sersic figure},
along with a histogram of $n$ indices. The profile fitting procedure
used in the NYU-VAGC
has a maximum of $n=5.90333$, resulting in the
artificial peak visible in this figure.
As with \cpetro, $n$ is affected by disk inclination, with
higher measured concentrations for more highly inclined galaxies.
The coloured histograms in Figure~\ref{sersic figure} show the distributions
of $n$ for each population identified in
Figure~\ref{cnorm-vs-ba} by their \cnorm\ values. It is clear
that the distinct populations of concentration identified by \cnorm\ also exist
when concentration is measured by the \sersic\ index,
and that the trimodality we detect is not an artefact
of the Petrosian concentration measure, but is an intrinsic
feature of the galaxy distribution.

\section{Properties of Intermediate Type Galaxies}
\label{intermediate properties}
In \S~\ref{trimodality} we discovered a third ``intermediate'' type
of galaxy that lies in a distinct region of morphological parameter
space from late-type disc galaxies and early-type elliptical galaxies.
In this section, we examine the properties of the galaxies that fall
into this classification.

\subsection{Global properties}
We begin by noting that the locus of intermediate type galaxies
in the \cpetro-$b/a$ plane has the same shape as the locus of late-type
galaxies that are clearly discs, and we may therefore infer that their
morphology is
at least partly disc-like. To estimate the degree of disciness in
these galaxies, we have plotted model bulge+disc systems for comparison
on Figure~\ref{cpetro-vs-ba} for three different $B/D$ ratios
(see Appendix~\ref{model-bd-systems} for details).
The $B/D=3$
curve lies along the locus of intermediate-type galaxies, suggesting
that these systems are 75\%\ bulge and 25\% disc;
note, however, that observed late-type disc galaxies are \textit{less}
concentrated than the $B/D=0$ model (see discussion in
Appendix~\ref{concentration-section}), and therefore our inferred $B/D$
for intermediate-type galaxies may be an underestimate.

We have split the galaxies into the regions denoted in
Figure~\ref{cnorm-vs-ba}: ``Disc'' galaxies with $\cnorm < 1.0$;
``Gap'' galaxies that lie between the disc peak and the intermediate
peak, $1.0 < \cnorm < 1.08$; ``Intermediate''-type galaxies with
$1.08 < \cnorm < 1.17$; and ``Elliptical'' galaxies that lie at higher
\cnorm\ and have higher axis ratios (see Figure~\ref{cnorm-vs-ba}
for the exact definition).
When we require samples that are unbiased in axis ratio, we extend
the elliptical sample to all axis ratios, although the galaxies with
$\cnorm > 1.17$ and lower axis ratios appear to have more in common
with intermediate-type galaxies than with true ellipticals.

\begin{figure*}
$\begin{array}{rcccc}
 & \cnorm=0.9~\textrm{(Disc)} & \cnorm=1.05~\textrm{(Gap)} &
 \cnorm=1.14~\textrm{(Intermediate)} & \cnorm=1.32~\textrm{(Elliptical)}\\[0.3cm]
\parbox[c]{1.35cm}{$b/a=0.9$} & \parbox[c]{115pt}{\includegraphics{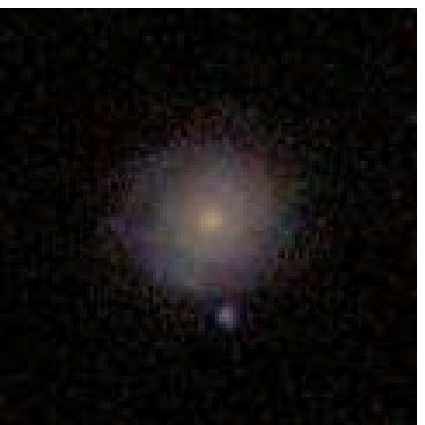}} &
  \parbox[c]{115pt}{\includegraphics{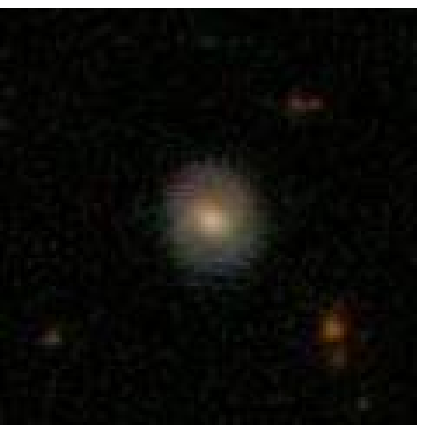}} &
  \parbox[c]{115pt}{\includegraphics{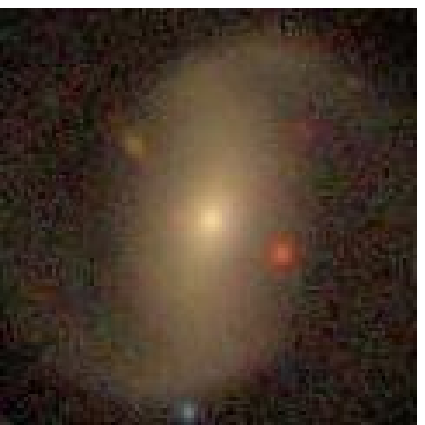}} &
  \parbox[c]{115pt}{\includegraphics{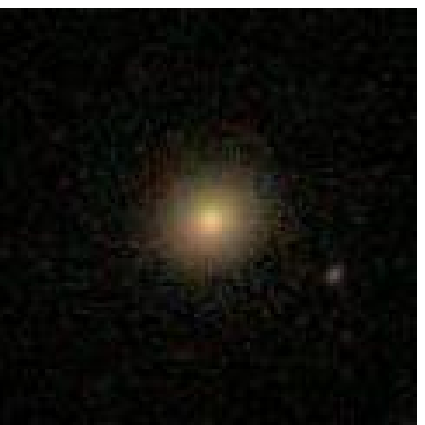}}\\[58pt]
\parbox[c]{1.35cm}{$b/a=0.7$} & \parbox[c]{115pt}{\includegraphics{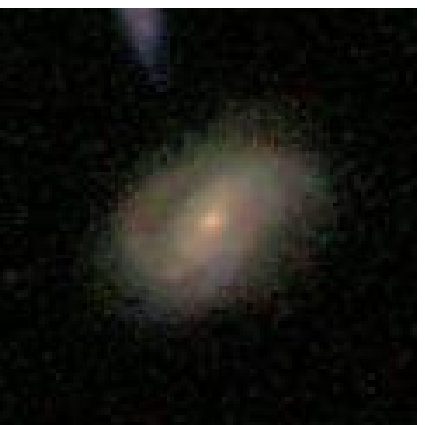}} &
  \parbox[c]{115pt}{\includegraphics{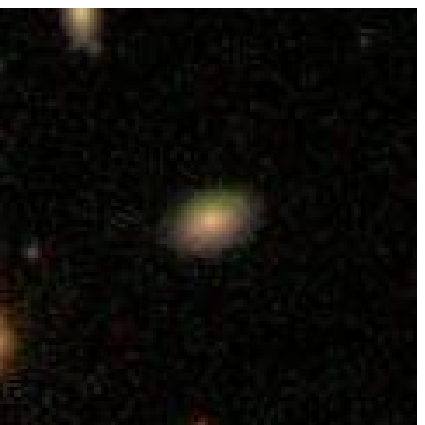}} &
  \parbox[c]{115pt}{\includegraphics{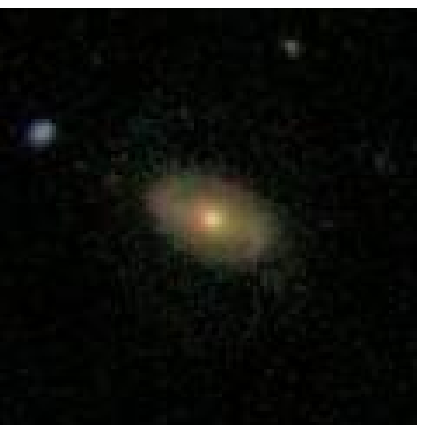}} &
  \parbox[c]{115pt}{\includegraphics{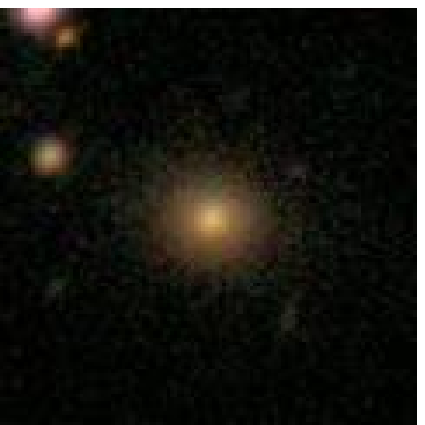}}\\[58pt]
\parbox[c]{1.35cm}{$b/a=0.5$} & \parbox[c]{115pt}{\includegraphics{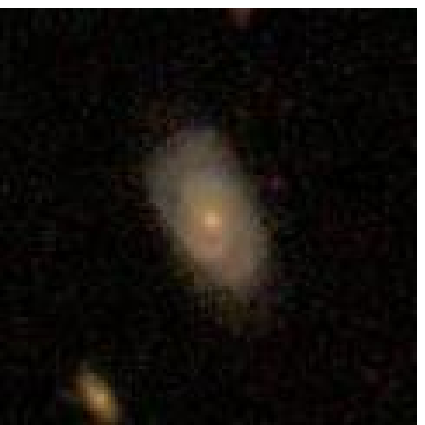}} &
  \parbox[c]{115pt}{\includegraphics{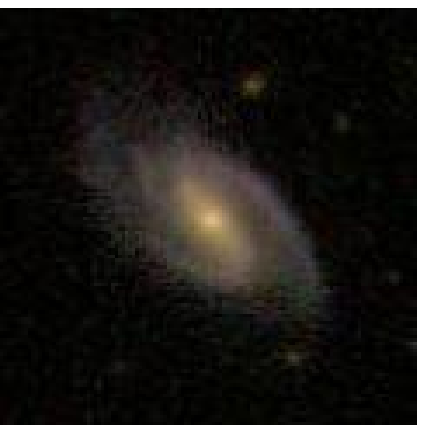}} &
  \parbox[c]{115pt}{\includegraphics{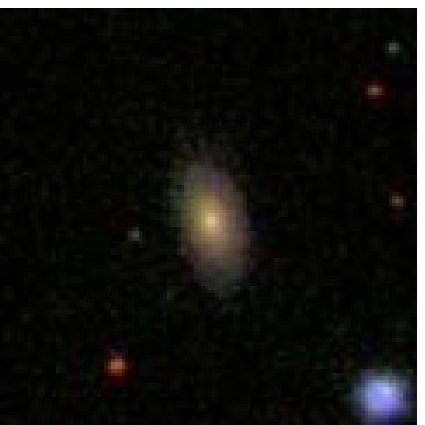}} &
  \parbox[c]{115pt}{\includegraphics{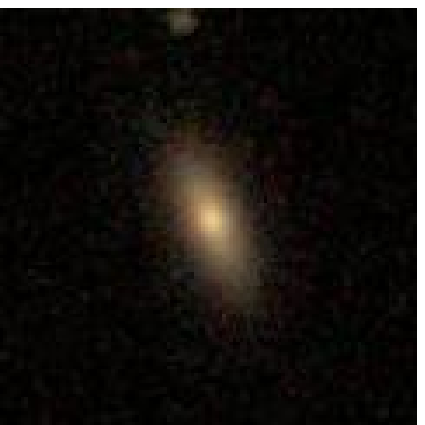}}\\[58pt]
\parbox[c]{1.35cm}{$b/a=0.3$} & \parbox[c]{115pt}{\includegraphics{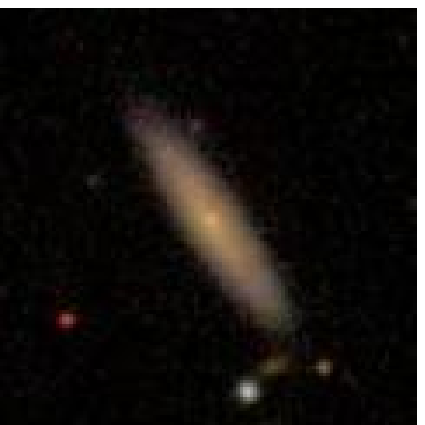}} &
  \parbox[c]{115pt}{\includegraphics{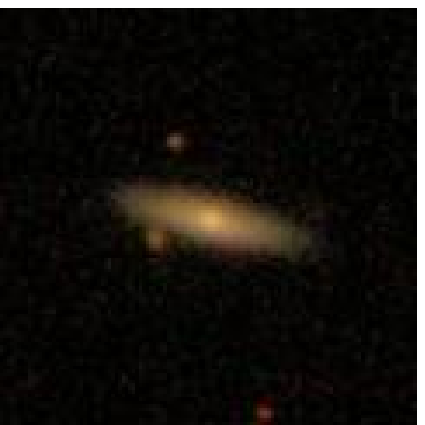}} &
  \parbox[c]{115pt}{\includegraphics{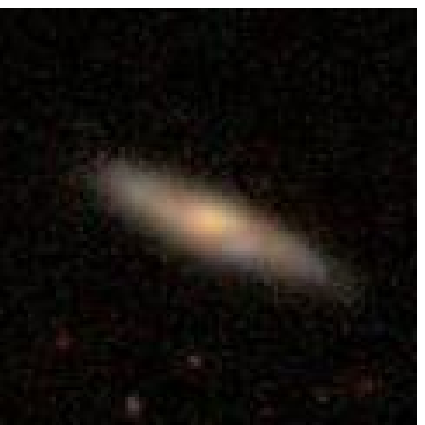}} &
  \parbox[c]{115pt}{\includegraphics{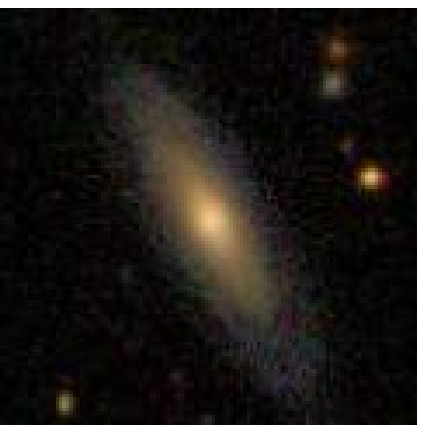}}\\
\end{array}$
\caption{\label{postage-stamps}%
Images of sample SDSS galaxies at $z<0.1$ with $-22 < \absmag < -21$
of various concentration and projected axis ratio. The columns correspond
to the disc, gap, intermediate and elliptical classifications from left
to right.}
\end{figure*}

\begin{figure*}
$\begin{array}{c@{\hspace{.25cm}}c}
\includegraphics[scale=0.5]{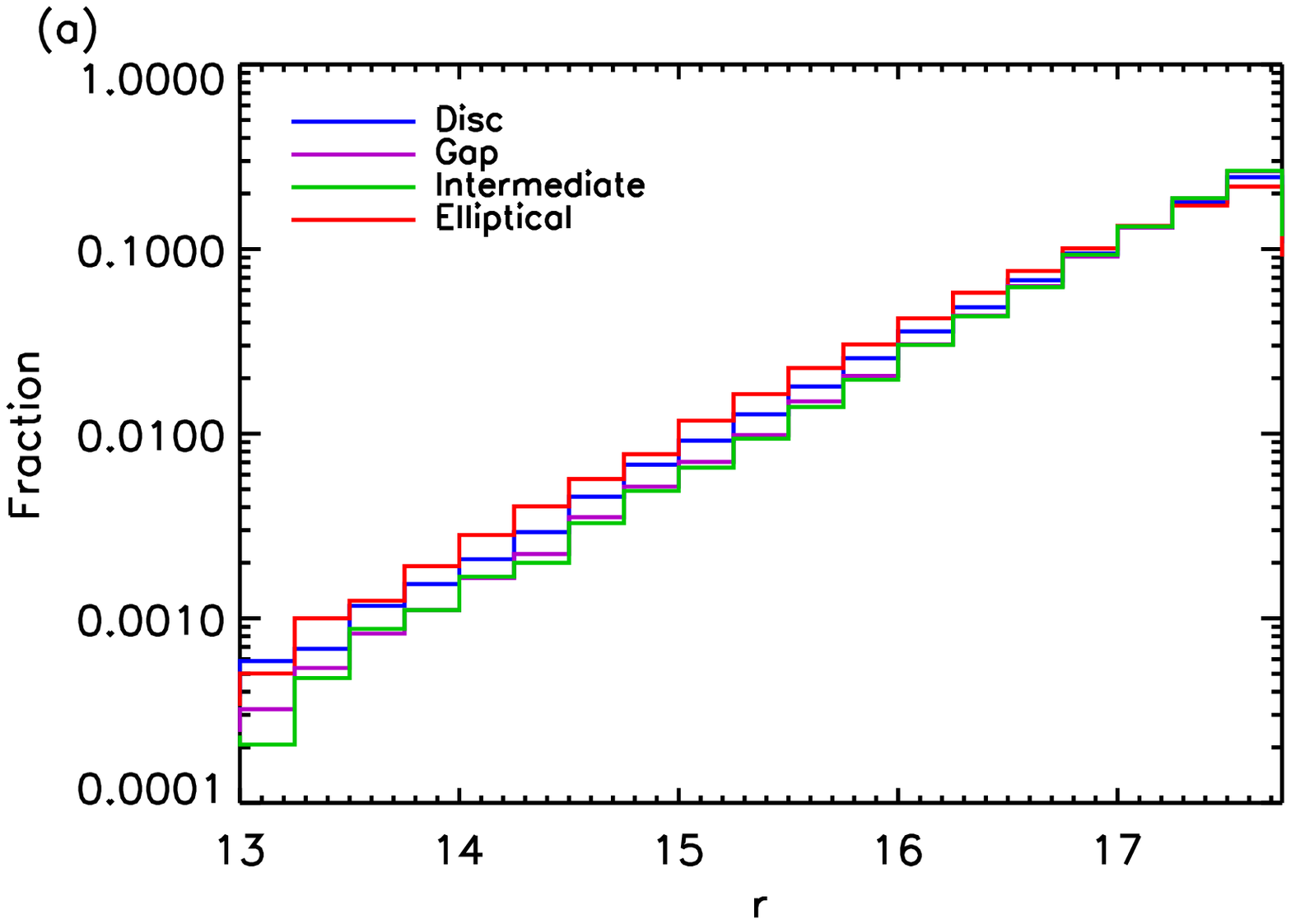} & 
\includegraphics[scale=0.5]{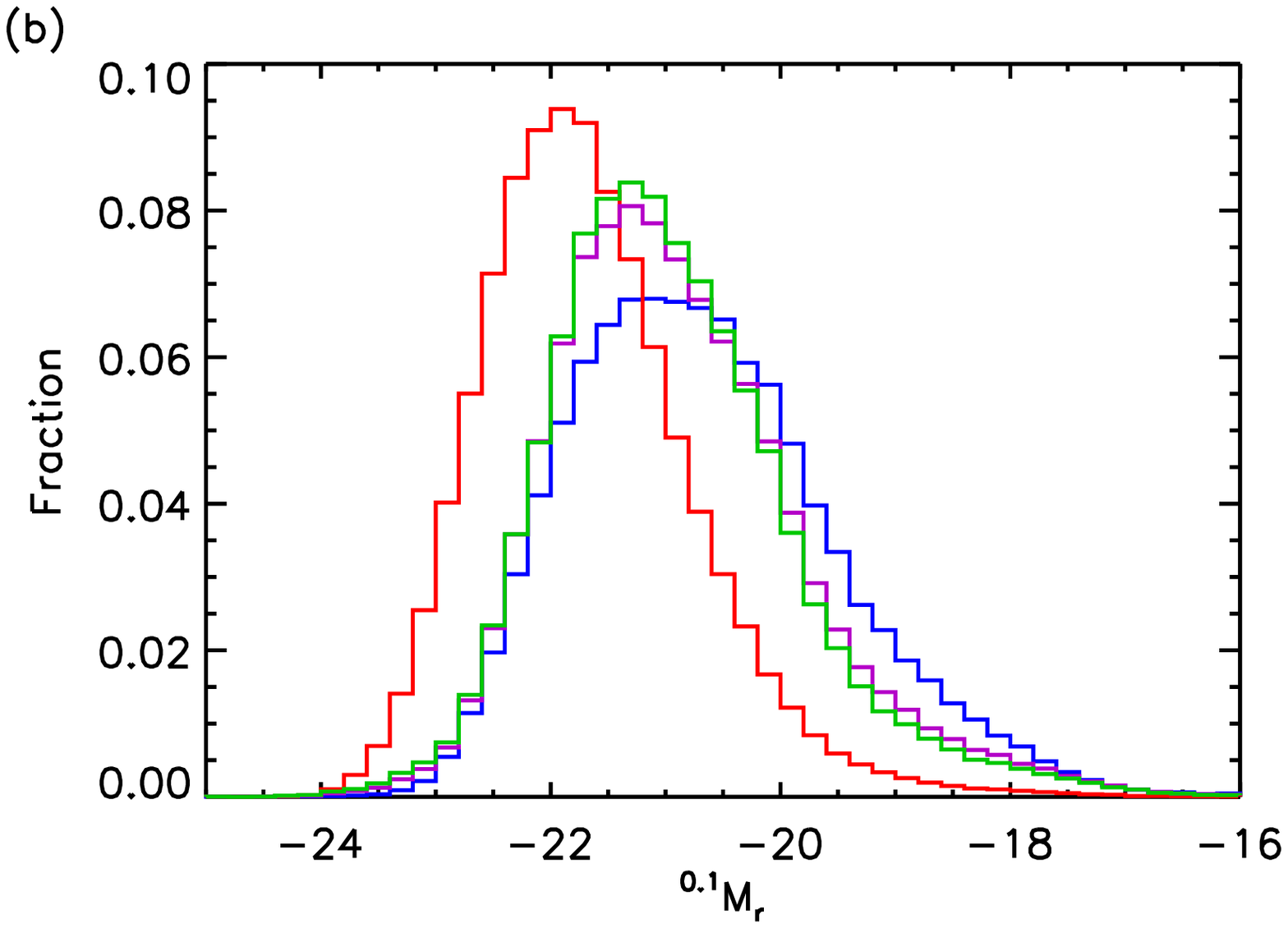}\\
\includegraphics[scale=0.5]{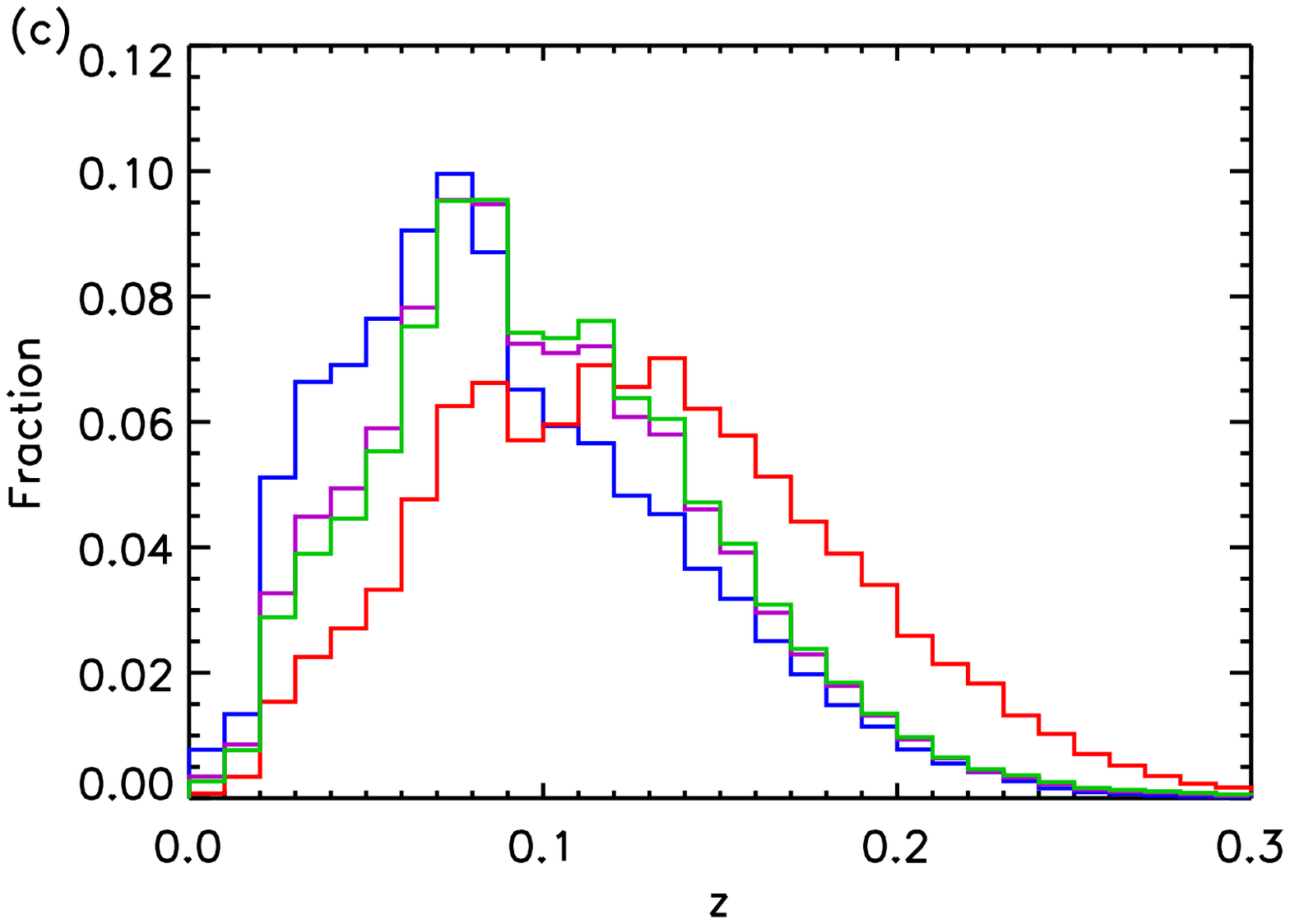} &
\includegraphics[scale=0.5]{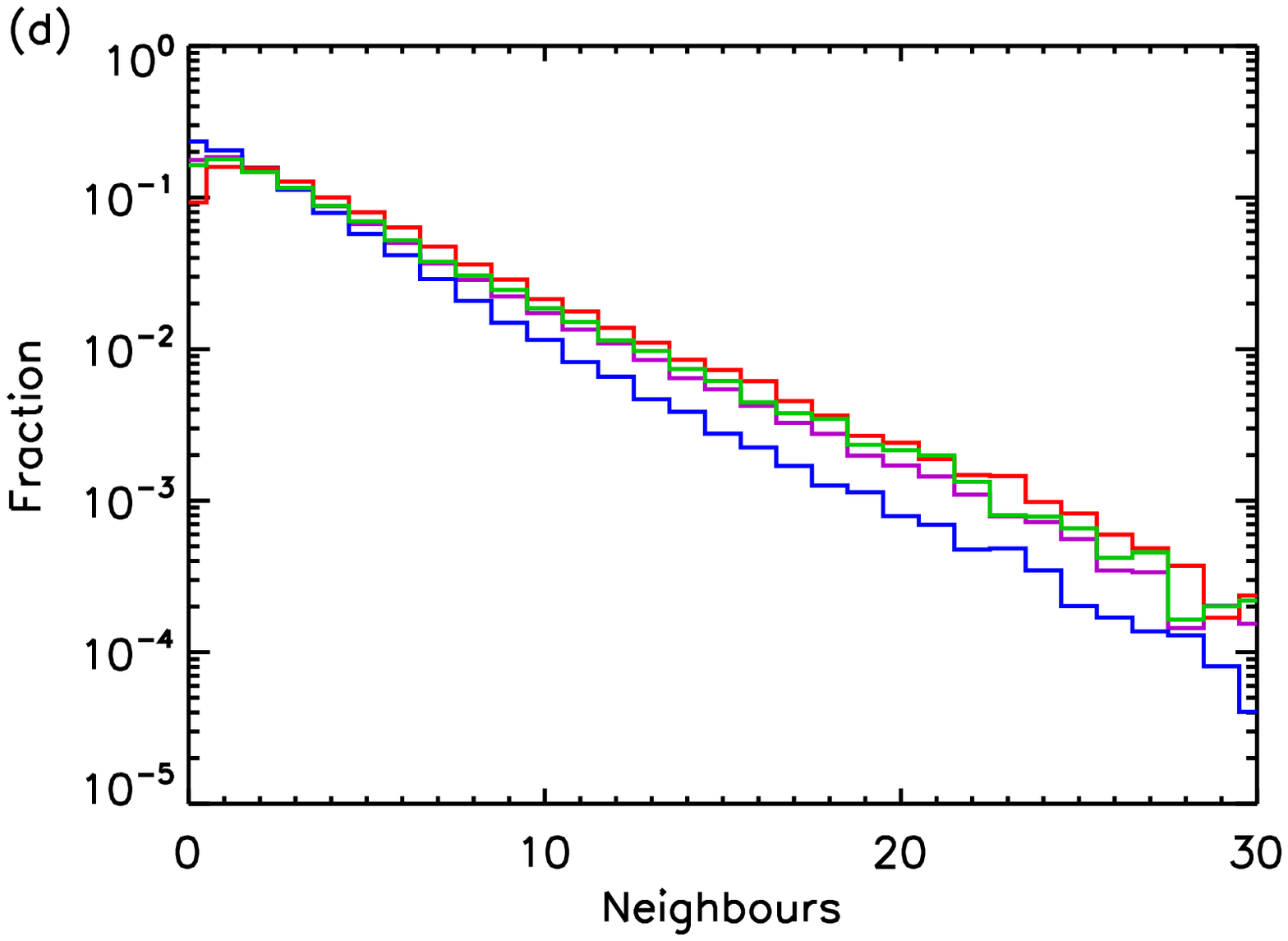}\\
\end{array}$
\caption{\label{magz}%
(a) Distribution of Petrosian r-band magnitudes of
disc (blue), elliptical (red), intermediate (green)
and gap (purple) galaxies in the SDSS Main Galaxy sample.
(b) Distribution of absolute dereddened ${}^{0.1}r$-band magnitudes,
\absmag.
Colours are as in panel (a).
(c) Distribution of redshifts. Colours are as in panel (a).
(d) Histogram of the number of $\absmag < -21$ neighbours within
a projected radius of $3$~Mpc and a velocity difference of $500~\kms$
around each $z<0.12$ galaxy. Colours are as in panel (a).}
\end{figure*}

In Figure~\ref{postage-stamps}, we show images of
SDSS galaxies that span the \cnorm-$b/a$ parameter space. They were
chosen to have typical absolute magnitudes ($-22 < \absmag < -21$%
\footnote{\absmag\ is the absolute
dereddened ${}^{0.1}r$-band magnitude, where
${}^{0.1}r$ is the $r$-band redshifted to $z=0.1$ and is used to minimize
errors due to uncertainties in the $k$-correction.};
see Figure~\ref{magz}) and to be close enough
to us for their morphology to
be apparent ($z<0.1$), but were otherwise chosen randomly. The ``disc'' galaxies
all appear to be late-type star-forming spiral galaxies. Both ``gap''
and ``intermediate'' galaxies appear redder than their disc counterparts.
Although they generally show evidence of discs, most are bulge-dominated and
would be classified as early-type spirals or possibly lenticulars.
Finally, all galaxies in the ``elliptical'' sample with large axis ratios
appear to be classical early-type passive elliptical galaxies, while those
with smaller axis ratios appear more like lenticulars.

Most properties of galaxies in the gap mirror those of intermediate
types. This is expected from examination of Figure~\ref{trimodal_fit},
which reveals that although
``gap'' galaxies are located at a local minimum in the
\cnorm\ distribution, their numbers
are dominated by the relatively wide intermediate peak.
We will therefore refer to ``gap'' and ``intermediate'' galaxies
together as ``intermediate'' types for the remainder of this
paper, except for the few cases where their properties differ.

The basic properties of the intermediate class of galaxies
are shown in Figure~\ref{magz}.
Panel (a) shows the observed $r$-band magnitude.
Intermediate-type galaxies in the SDSS Main Galaxy sample are on average
fainter than disc galaxies, which are themselves fainter than elliptical
galaxies.
Panel (b) shows the absolute magnitude distribution.
The absolute magnitudes of intermediate-type
galaxies are similar to those of disc galaxies at the luminous end, but
there are fewer intermediate-type galaxies in the low-luminosity
tail; in contrast, the distribution of
elliptical galaxies peaks $0.6$ magnitudes brighter.
The redshift distribution is shown in panel (c). Although the shape
of the $z$ distribution for intermediate-type galaxies is more similar
to that of late-type discs than of early-type ellipticals, the peak of
the distribution is shifted to higher redshifts, which is responsible
for the fainter apparent magnitudes seen in panel (a).

In panel (d), we attempt to quantify the density of the environment
surrounding each galaxy in order to investigate how the three galaxy
types participate in the well-known morphology-density relation
\citep[e.g.][]{dressler80}.
We quantify the local environment by the number of neighbouring galaxies
at $\absmag < -21$ within a projected radius of $3$~Mpc
and with observed redshifts within $500$~\kms\ of
the galaxy in question. We only use galaxies at $z<0.12$ in this figure,
for which the sample of $\absmag < -21$ neighbours is complete to the SDSS
Main Galaxy magnitude limit.
We recover the morphology-density relation:
galaxies surrounded by more than $2$ bright neighbours are far more likely
to be elliptical galaxies than to be disc galaxies, while galaxies
surrounded by less than $2$ bright neighbours are far more likely to
be discs. The environments of intermediate-type galaxies are almost
exactly intermediate between those of discs and ellipticals:
the fraction of galaxies in low density regions,
i.e.  with less than $2$ bright neighbours,
is $44\%$, $34\%$ and $25\%$ for ``disc'',
``intermediate'' and ``elliptical'' galaxies respectively.

\subsection{Number and luminosity density}
\begin{figure}
\includegraphics[scale=0.5]{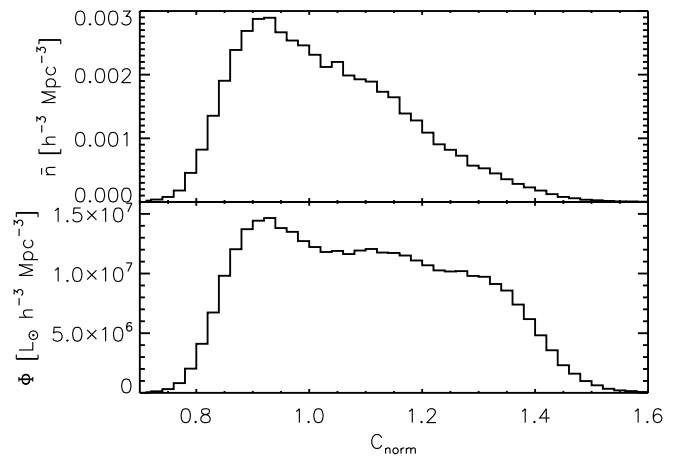}
\caption{\label{cndens}%
Comoving number density (top) and ${}^{0.1}r$-band luminosity density
(bottom) of SDSS galaxies with $\absmag < -17$.}
\end{figure}

In order to determine the fraction of the total galaxy population
represented by the intermediate-type galaxies,
we have plotted the comoving number
density and luminosity density of SDSS galaxies as a function of
\cnorm\ in Figure~\ref{cndens},
correcting for the effective surveyed volume $V_{\mathrm{max}}$
for each galaxy.
When calculating the volume densities, we have restricted the survey to
$0.01 < z < 0.2$, $\absmag < -17$, and $r < 17.7$
in order to minimize
statistical fluctuations in the $1/V_{\mathrm{max}}$ term. By number,
late-type disc galaxies appear most prominent,
although a large bump at intermediate
concentrations is visible. When weighted by luminosity,
the trimodality of the galaxy
population is very apparent.
We have performed unconstrained trimodal fits to both distributions
and listed the parameters in Table~\ref{rmix fit table}.
The location and width of each peak match those found
from fitting the raw number counts very well, indicating that
the fits are identifying the same populations. We conclude from these
fits that intermediate-galaxies have a comoving number density
of $0.03~h^{-3}~\mathrm{Mpc^3}$
and a comoving ${}^{0.1}r$-band luminosity density of
$2 \times 10^8~L_{\sun}~h^{-3}~\mathrm{Mpc^3}$, representing $\sim 60\%$
of the total number density and $\sim 50\%$ of the total luminosity density of
$\absmag < -17$ galaxies.

\subsection{Colours, spectra and apparent shapes}
\begin{figure}
\includegraphics[scale=0.5]{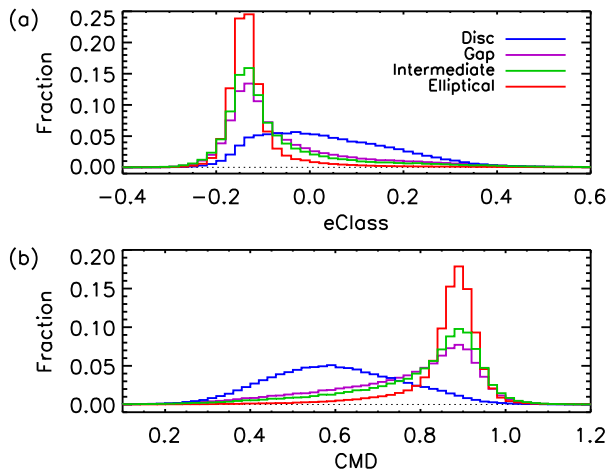}
\caption{\label{sfh}%
(a) Distribution of the spectroscopic PCA eClass parameter
for disc (blue), elliptical (red), intermediate (green) and gap (purple)
galaxies in SDSS.
(b) Distribution of the CMD location parameter (i.e. colour relative to
the red sequence) for SDSS galaxies. Colours are as in panel (a).}
\end{figure}

In Figure~\ref{sfh}, we show how observables that
are sensitive to star formation history vary among galaxy types.
We quantify the star formation history using two properties:
the spectroscopic PCA eClass parameter, which is positive for star-forming
galaxies and negative for quiescent galaxies, and the
location of the galaxy on the colour-magnitude diagram (\CMD), which we
quantify as
\begin{equation}
\CMD = {}^{0.1}(g-r) + 0.0325\left(\absmag + 19.7745\right),
\end{equation}
i.e. it is the ${}^{0.1}(g-r)$ colour after taking out the slope of the
colour-magnitude relation for SDSS galaxies, normalized at
$\absmag - 5\log h = -19$.
Late-type blue galaxies
have $\CMD \la 0.78$ while early-type red galaxies have $\CMD \ga 0.78$.
The distribution of star formation histories among intermediate-type
galaxies does not appear to be the sum of disc and elliptical
contributions, which would result in two distinct peaks (particularly
in the CMD distribution), but instead is dominated by a distinct
peak of quiescent red galaxies with a tail of residual blue star-forming
galaxies.

\begin{figure}
\includegraphics[scale=0.5]{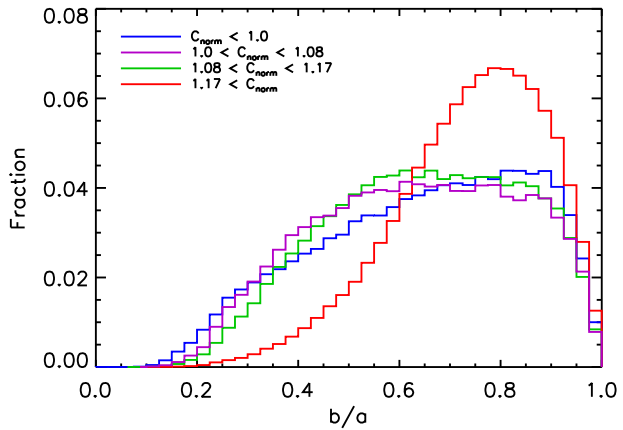}
\caption{\label{ba-hist}%
Histogram of isophotal $b/a$ axis ratios for disc (blue), elliptical (red),
intermediate (green) and gap (purple) galaxies in SDSS. In this plot
we include all galaxies with $\cnorm>1.17$ in the elliptical sample
regardless of axis ratio.}
\end{figure}

In Figure~\ref{ba-hist}, we plot the distribution of apparent axis ratios
$b/a$ for galaxies of different concentration \cnorm. For this figure,
we include all galaxies with $\cnorm > 1.17$ in the ``elliptical'' bin
in order to have a sample that is unbiased in $b/a$. The distribution of
elliptical galaxy shapes is strongly peaked at $b/a=0.8$, and very
few galaxies have small axis ratios; as seen in Figure~\ref{postage-stamps},
those few very flattened galaxies appear more similar to
``intermediate''-type galaxies than to circular elliptical galaxies.
The disc galaxies also peak at high $b/a$, but with a very broad
distribution that includes many very flattened systems, as expected for
their intrinsically disc-like morphology.

On the other hand, the apparent axis ratios of ``intermediate'' galaxies
is peculiar. It peaks at $b/a\approx 0.6$, with a decline to both
lower and higher apparent axis ratios, and
the peak at low $b/a$ would be even further enhanced if the
low-$b/a$ ``elliptical'' galaxies were considered as intermediate types.
If the morphology of intermediate galaxies is relatively
disc-like, as implied by the distribution of intermediate-type galaxies in
the \cpetro-$b/a$ plane, then why is there a deficit of apparently
round galaxies? If the morphology is a combination of $25\%$ flattened disc
with $75\%$ almost spherical bulge, as suggested by the dashed line
in Figure~\ref{cpetro-vs-ba}, how can the combination appear more flattened
than either pure discs or pure bulges?

We suggest five possible explanations for the unusual distribution
of apparent axis ratios of intermediate-type galaxies:
\begin{enumerate}
 \item The discs may be elliptical, and the decline as $b/a \rightarrow 1$
reflects the deficit of intrinsically axisymmetric galaxies. Indeed, the
dark matter haloes that host galaxies are expected to be triaxial
\citep[and references therein]{allgood-etal06}, and therefore galactic
discs that form in these haloes should be elliptical.
However the required typical disc ellipticity of $\sim 0.4$
is much higher than the median disc ellipticity of $0.13$ that \citet{ryden04}
inferred from $r$-band isophotal axis ratios or than the ellipticities
that can be generated in self-consistent massive discs by triaxial dark
matter haloes \citep{bailin-etal07-triaxialdisk}.
 \item The galaxies may be intrinsically prolate objects, and therefore
appear elongated from most viewing angles. We consider this
unlikely both because the locus of intermediate-type galaxies in the
\cpetro-$b/a$ plane has the same shape as that of the late-type galaxies that
are known to be discs, and because early-type galaxies with high
ellipticities have predominantly discy isophotes \citep{hao-etal06}.
 \item \citet{ar02} found that red disc galaxies in SDSS have higher
apparent ellipticities than blue disc galaxies
simply due to the combination of dust and inclination
effects, which causes edge-on disc galaxies to appear redder. As our
intermediate galaxies are redder than our disc galaxies
(see Figure~\ref{sfh}) and show evidence of dust (see below),
they could simply be the analog of \citet{ar02}'s dusty edge-on discs.
However, we know that the disc and intermediate galaxies are intrinsically
different for two reasons: (1) the effect of dust, which is centrally
concentrated, would be to \textit{decrease} the measured concentration when
discs
are viewed edge-on, while our intermediate galaxies have \textit{higher}
concentrations than our disc galaxies; and (2) the gap between the disc
and intermediate galaxies in the \cnorm-$b/a$ plane persists even to the
most face-on $b/a \sim 1$ galaxies, where dust can no longer be important.
We therefore discount this explanation.
 \item Intermediate-type galaxies with face-on discs may appear \textit{more}
concentrated than their edge-on counterparts, and therefore have large
enough \cnorm\ to fall into the elliptical classification. There are
two potential effects that could cause this:
 \begin{enumerate}
   \item The measured concentration of bulge+disc systems is not a monotonic
  function of bulge-to-disc ratio; as seen from the $B/D=10$ line
  in Figure~\ref{cpetro-vs-ba}, the addition of a small face-on disc
  to a spheroidal galaxy increases the measured concentration over
  that of a small edge-on disc or even of a pure spheroid!
  The reason for this is investigated in Appendix~\ref{concentration-section}.
  A consequence is that lines of constant $B/D$ in the \cnorm-$b/a$ plane
  can be tilted to higher \cnorm\ at higher $b/a$.
  However, we do not find a single value of $B/D$ that can both explain
  galaxies on the intermediate-type locus and in the elliptical cloud.
   \item The face-on intermediate galaxies may contain a luminous feature
  in the central region that is obscured by dust when seen edge-on.
  As noted below, we see evidence
  for dust-obscured central star formation in the high-eccentricity
  intermediate-type population; such star formation is likely to contribute
  a non-negligible amount of flux in the inner regions of the galaxy,
  which would increase the measured concentration when it is visible.
 \end{enumerate}
Evidence for an explanation of this form comes from the apparent bridge between
the elliptical and intermediate groups at $\cnorm \approx 1.18$ and
$b/a \approx 0.6$.
\end{enumerate}
Our favoured explanation, which is consistent with all the observations,
is the last one: face-on intermediate-type galaxies appear to be ellipticals
in the \cpetro-$b/a$ plane due to central star formation that increases the
measured concentration.

\begin{figure}
\includegraphics[scale=0.5]{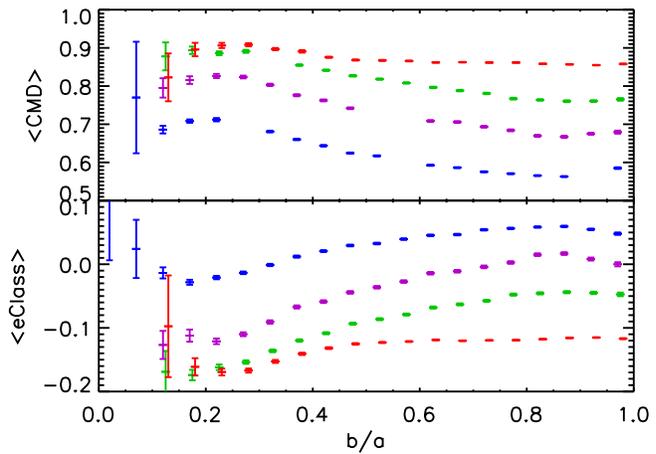}
\caption{\label{sfh-vs-ba}%
(Top) Mean CMD location parameter for disc (blue), elliptical (red),
intermediate (green) and gap (purple) galaxies in SDSS as a function
of their isophotal $b/a$ axis ratio.
(Bottom) Mean spectroscopic PCA eClass parameter for SDSS galaxies as
a function of their isophotal $b/a$ axis ratio. Colours are as in the
top panel.
Note that in this Figure, the elliptical sample includes all galaxies with
$\cnorm > 1.17$ regardless of axis ratio.}
\end{figure}

To further investigate how the properties of galaxies depend on the axis
ratio, we have plotted in Figure~\ref{sfh-vs-ba} the mean \CMD\ and eClass
parameters as a function of apparent axis ratio.
As in Figure~\ref{ba-hist}, the ``elliptical'' class here includes all
galaxies with $\cnorm > 1.17$ regardless of $b/a$.
We do not expect the intrinsic star formation properties of galaxies of
a given morphological type to depend strongly on the intrinsic axis ratios,
and therefore any variation we see is likely due to obscuration effects.
Indeed, we find that the properties of
elliptical galaxies, which do not contain
much dust, are almost independent of apparent axis ratio for $b/a>0.4$.
Among the more flattened ellipticals, there is a trend, but it is almost
identical to that of
the intermediate galaxies, and we believe that the galaxies in this
region of parameter space are more likely to be intermediate galaxies
with unusually high concentrations than elliptical galaxies with
unusually small axis ratios.
Disc galaxies show a significant tendency to appear redder and to have
less apparent star formation when viewed edge-on than face-on. This is
most easily interpreted as due to dust obscuration of star forming
regions. The apparent star formation
properties of intermediate galaxies show a perfectly analogous 
trend with apparent
axis ratio as disc galaxies: they appear significantly redder and more
quiescent when viewed edge-on. However, at a given axis ratio the
intermediate galaxies are $\sim 2$ magnitudes redder and have
an eClass that is $\sim 0.15$ lower than the equivalent disc galaxies.
Among the most edge-on galaxies, the redness and apparent
quiescence of intermediate-type galaxies
even surpass those of the ellipticals. Therefore, many intermediate-type
galaxies must contain dust that obscures star formation when they are
viewed edge-on. It is interesting to note that this is the only figure
(and, to a much lesser degree, Figure~\ref{sfh})
in which we find a significant difference between the properties of
``gap'' and ``intermediate'' galaxies: the colours and spectra of
``gap'' galaxies of a given apparent axis ratio lie exactly between those
of the equivalent ``disc'' and ``intermediate'' galaxies.

\subsection{Sizes}
\begin{figure}
\includegraphics[scale=0.5]{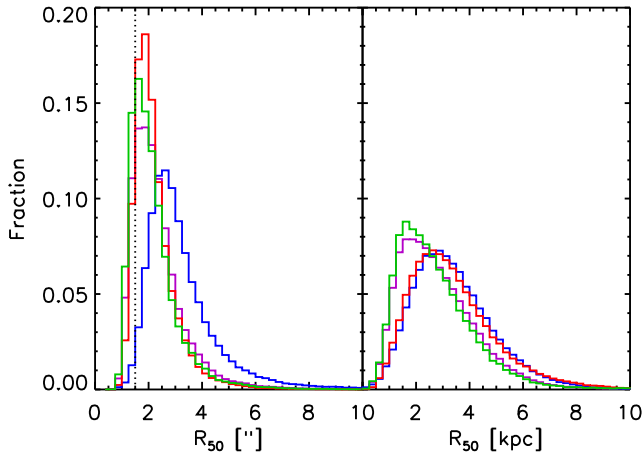}
\caption{\label{size-plot}%
Histograms of the
Petrosian half-light radii $R_{50}$, for disc (blue), elliptical (red),
intermediate (green) and gap (purple) SDSS galaxies,
measured in angular units (left) and
physical units (right).
The seeing FWHM of $1.5\arcsec$ is
denoted with the vertical dotted line in the left panel.}
\end{figure}

In Figure~\ref{size-plot}, we plot the Petrosian half-light radii $R_{50}$ of
our galaxies, both in angular size and in physical kpc.
A clear feature in both panels is that the intermediate class of galaxies
are smaller than the other classes of galaxies.

The typical seeing FWHM in the SDSS images is $1.5\arcsec$,
so a point source would have a measured half-light radius of
$1.2\arcsec$, and $R_{50}$ values that are in this regime may
have been artificially inflated by seeing.
As the median $R_{50}$ of the intermediate-type galaxies is only $2.0\arcsec$,
perhaps the concentration of these galaxies
is subject to resolution effects and the presence of a peak at
intermediate concentrations is merely an artefact.

\begin{figure}
\includegraphics[scale=0.5]{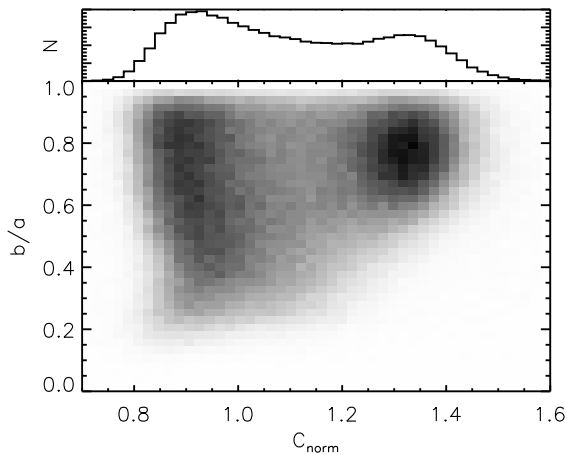}
\caption{\label{r50-cnorm-ba}%
As in Figure~\ref{cnorm-vs-ba}, but only including galaxies with
$R_{50} > 1.75\arcsec$.}
\end{figure}

In Figure~\ref{r50-cnorm-ba},
we have reconstructed the \cnorm-$b/a$ distribution
and \cnorm\ histogram
of Figure~\ref{cnorm-vs-ba} using only the most
well-resolved galaxies, those with $R_{50} > 1.75\arcsec$.
The intermediate peak in the histogram has indeed disappeared, as has the
strong vertical feature at $\cnorm \sim 1.14$, $b/a \la 0.6$
However, there is still 
a significant population at $1.08 < \cnorm < 1.17$ and $b/a < 0.65$,
and the very bottom-heavy distribution of $b/a$ among
the galaxies at intermediate concentrations is still evident.
We also note that among these well-resolved galaxies,
the elliptical galaxy cloud is much smaller and has moved to higher
\cnorm.

There are two possible explanations for the $R_{50}$ distribution
of intermediate galaxies:
\begin{enumerate}
 \item The trimodality is an artefact of seeing.
Specifically, some population of galaxies from an intrinsically
bimodal distribution artificially congregates in the
``intermediate'' region of parameter space when poorly resolved.
 \item The intrinsic sizes of intermediate-type galaxies are smaller
than elliptical and disc galaxies, and they therefore happen to typically
be poorly-resolved given their redshift distribution.
\end{enumerate}

We consider the first explanation unlikely for the following reasons.
Firstly, the difference between the size distributions of intermediate
galaxies vs. of discs and ellipticals is most prominent
when plotted in physical units, rather than angular units.
Secondly, a trimodal fit to the \cnorm\ distribution of just
the most well-resolved galaxies is strongly preferred over a bimodal
fit, with an intermediate peak located at the same mean
\cnorm\ and with the same width as the fit to the full sample
(see Table~\ref{rmix fit table}).
Thirdly, seeing spreads the central light of a galaxy out, and therefore
the net effect should be to decrease \cpetro. This is indeed what we
see among the elliptical galaxies in Figure~\ref{r50-cnorm-ba}: the elimination
of the poorly-resolved galaxies increases the typical concentration.
Seeing also circularizes the light profile, but at a much smaller
radius than the isophotal radius for most of these galaxies. Therefore,
the effect of seeing is to move galaxies to the left (and possibly upwards)
on Figure~\ref{cnorm-vs-ba}.
In order for a population of galaxies to be artificially
moved onto the intermediate-type locus by seeing, that population must lie
to the right of (and possibly down from) the intermediate galaxies.
However, this
part of parameter space, at high \cnorm\ and low $b/a$, is among
the most sparsely populated in Figures~\ref{cnorm-vs-ba} and 
\ref{r50-cnorm-ba},
making this explanation very unlikely.

Since \cpetro\ does not always behave as expected
(see Appendix~\ref{concentration-section}),
another possibility is that for these galaxies, \cpetro\ is actually
\textit{increased} by the effects of seeing, and the intermediate
region of parameter space is populated by poorly-resolved disc
galaxies. However, this explanation is also untenable: intermediate-type
galaxies are red and have passive spectra while disc galaxies are
blue and have star forming spectra.
Moreover, the distribution of axis ratios
among intermediate-type galaxies is dominated by low values of $b/a$
while those of disc galaxies are dominated by high $b/a$, and seeing
cannot decrease $b/a$.

While the second explanation is uncomfortable in that it proposes that
a new class of galaxies identified morphologically
are coincidentally not well-resolved, it is
the only explanation that is consistent
with the other properties of these galaxies, and we therefore conclude
that it is correct.

\section{Discussion}
\label{discussion}
The key question prompted by our results is why there is
an overdensity of galaxies at $B/D \sim 3$ with unique properties.
These galaxies appear
to be a combination of early-type spiral galaxies and S0/lenticular
galaxies; does the existence of
this overdensity imply that early spirals and lenticulars
have a unique formation mechanism?

The most commonly proposed scenario for forming lenticulars is
that they are the remnants of later-type disc galaxies whose discs
fade (and therefore whose $B/D$ ratios increase)
when they run out of fuel for star formation
due to rapid gas loss after they enter the cluster environment,
either through ram pressure stripping \citep{gg72}
or strangulation \citep{ltc80}.
The morphological transformation may also be augmented
by the effects of galaxy-galaxy interactions within the denser
environment \citep{bekki98}.
Some properties of the intermediate-type galaxies in our sample
are consistent with this scenario. In particular, the small
physical scale sizes of intermediate-type galaxies are
precisely what would be expected if the luminosity of the
extended disc declines relative to that of the concentrated bulge.
On the other hand, if a cluster environment is required
to produce intermediate-type galaxies, then they should
be found exclusively in clusters and be even more biased towards lying
in clusters than ellipticals,
while we find that $34\%$ of intermediate-type galaxies
are in low-density environments, significantly more than the
$25\%$ of elliptical galaxies found in these regions.
Finally,
if intermediate-type galaxies are produced by faded disc-type galaxies,
then the bright end of the intermediate-type sample would be fainter
than the bright end of the disc-type sample,
while we find that
the brightest intermediate-type galaxies are, if anything,
brighter than the brightest discs.
However, if the bright end of the
disc sample has already been depopulated by this
process at $z=0$ then the observed luminosity distributions may be consistent
with the disc-fading scenario.

Interestingly, \citet{bendo-etal07} find a qualitative difference
between the compactness of the $24\micron$ dust emission,
a good tracer of the interstellar medium (ISM), in early-type spirals
(S0--Sab) and late-type spirals (Sc and later): a large fraction
of the early-type spirals have a centrally-concentrated ISM, while
virtually all of the late-type spirals have extended ISM distributions.
This dichotomy they see between the ISM concentration of
the two classes of disc galaxies
mirrors the dichotomy we see
between the concentration of the stellar light between late-type disc
galaxies and the more bulge-dominated intermediate-type galaxies.

One further consequence of our results is that a non-negligible
fraction of galaxies that have been identified as early-type galaxies
in studies that classify galaxies using colours, spectra,
or even \cpetro, are actually members of a distinct intermediate
class. Even in this study, it may be that many of the face-on
intermediate galaxies are misidentified as elliptical galaxies.

\section{Conclusions}
\label{conclusions}
We have analysed the locations of SDSS galaxies in the Petrosian concentration
\cpetro\ vs. isophotal axis ratio $b/a$ plane. Disc galaxies occupy a
slanted curve in this plane due to inclination effects: more edge-on
galaxies have higher \cpetro. We have corrected for this and defined
a new inclination-independent concentration index \cnorm.
Unlike most galaxy properties (including \cpetro), which are
distributed bimodally, the distribution of \cnorm\ is
\textit{tri}modal, with a third peak intermediate between the
low-concentration disc galaxies and the high-concentration
elliptical galaxies.

We have studied the properties of this newly-identified intermediate class
of galaxy, which represents $\sim 60\%$ of the total number density and
$\sim 50\%$ of the luminosity density of $\absmag < -17$ galaxies
detectable in SDSS.
Most intermediate-type galaxies have colours and spectra typical of
early-type galaxies, although a non-negligible number are blue and
star forming. The location of these galaxies on the \cpetro-$b/a$
plane indicates that they have discs. The concentrations are indicative
of typical bulge-to-disc ratios of $3$, although this may be an
underestimate. The colours and spectra of intermediate-type galaxies
vary with inclination (edge-on galaxies are redder and appear to be
more passive), indicative of dust.
The fraction of blue star-forming galaxies among the intermediate class
may be underestimated due to dust obscuration of star-forming regions.
The typical absolute magnitudes and
half-light radii of the SDSS intermediate-type galaxies are $-21$ and
$2.4~\mathrm{kpc}$ respectively. The physical sizes of the intermediate-type
galaxies are significantly smaller than
those of both ellipticals and discs.

Intermediate-type galaxies have much higher apparent ellipticities than either
disc or elliptical galaxies. Our preferred explanation is that these
galaxies contain residual central star formation that is obscured by
dust when viewed edge-on; therefore, the measured concentration is
higher when viewed face-on and these galaxies merge into the region
of parameter space occupied by high-concentration ellipticals.

Associating our intermediate class of galaxies with lenticulars
and early-type spirals,
we propose that the existence of a distinct peak in parameter space
implies that there is a particular formation mechanism responsible
for the creation of the earliest-type spiral galaxies.

\section*{Ackowledgments}
We thank Greg Rudnick and Dennis Zaritsky for useful conversations,
and the anonymous referee of a previous paper, whose comments
sparked this line of inquiry.

    Funding for the Sloan Digital Sky Survey (SDSS) and SDSS-II has been provided by the Alfred P. Sloan Foundation, the Participating Institutions, the National Science Foundation, the U.S. Department of Energy, the National Aeronautics and Space Administration, the Japanese Monbukagakusho, and the Max Planck Society, and the Higher Education Funding Council for England. The SDSS Web site is http://www.sdss.org/.

    The SDSS is managed by the Astrophysical Research Consortium (ARC) for the Participating Institutions. The Participating Institutions are the American Museum of Natural History, Astrophysical Institute Potsdam, University of Basel, University of Cambridge, Case Western Reserve University, The University of Chicago, Drexel University, Fermilab, the Institute for Advanced Study, the Japan Participation Group, The Johns Hopkins University, the Joint Institute for Nuclear Astrophysics, the Kavli Institute for Particle Astrophysics and Cosmology, the Korean Scientist Group, the Chinese Academy of Sciences (LAMOST), Los Alamos National Laboratory, the Max-Planck-Institute for Astronomy (MPIA), the Max-Planck-Institute for Astrophysics (MPA), New Mexico State University, Ohio State University, University of Pittsburgh, University of Portsmouth, Princeton University, the United States Naval Observatory, and the University of Washington.

\bibliography{../../masterref.bib}

\appendix

\section{Models of Bulge+Disc Systems}
\label{model-bd-systems}
We have constructed models of bulge+disc systems in order to
determine the Petrosian concentration \cpetro\ and isophotal axis
ratio $b/a$ that would be measured when a given system is viewed
at an inclination $i$, where $i=0\degr$ is face-on.
To do this, we first calculate the galactic density on a fine 3D grid:
\begin{equation}
\rho(x,y,z) = \rho^{\mathrm{disc}}(x,y,z) + \rho^{\mathrm{bulge}}(x,y,z),
\end{equation}
where
\begin{equation}
\rho^{\mathrm{disc}} = \rho^{\mathrm{disc}}_0 \exp\left( -R/R_d
  \right) \sech^2 \left( z' / z_h \right),
\end{equation}
\begin{equation}
\rho^{\mathrm{bulge}} = \frac{\rho^{\mathrm{bulge}}_0}{r\left(r+a\right)^3},
\end{equation}
and the transformation from the Cartesian observed coordinates to
the cylindrical/spherical galactic coordinates are
\begin{equation}
z'(x,y,z) = -y \sin i + z \cos i
\end{equation}
\begin{equation}
R(x,y,z) = \sqrt{x^2 + (y \cos i + z \sin i)^2}
\end{equation}
\begin{equation}
r(x,y,z) = \sqrt{x^2 + y^2 + z^2}.
\end{equation}
The \citet{hernquist90} bulge
that we use has a projected surface density profile very
similar to the empirical \citet{devaucouleurs48} $R^{1/4}$ profile,
with $a=R_e/1.8153$ for effective radius $R_e$,
but is much simpler than the 3D deprojection of the $R^{1/4}$ law.
The normalization constants $\rho^{\mathrm{disc}}_0$ and
$\rho^{\mathrm{bulge}}_0$ are chosen so that the ratio of the total
mass of each component is equal to the desired bulge-to-disc ratio, $B/D$.  
We then project the 3D distribution down to 2D by summing over the $z$
axis. We set down annuli $0.05 R_d$ wide with inner radii ranging from $0$ to
$10 R_d$, and calculate the total flux in each annulus along with the mean flux
per unit area within the annulus.
The Petrosian radius $R_{\mathrm{Petro}}$ is defined to be
the radius at which the ratio of the local surface brightness in the
annulus to the mean enclosed surface brightness (the Petrosian ratio)
equals $0.2$.
The Petrosian flux is the enclosed flux within $2 R_{\mathrm{Petro}}$ and
the Petrosian concentration \cpetro\ is defined to be the ratio
of $R_{90}$ and $R_{50}$, the radii
enclosing $90\%$ and $50\%$ of the Petrosian flux respectively:
\begin{equation}
\cpetro \equiv \frac{R_{90}}{R_{50}}.
\end{equation}

Finally, we calculate the isophotal $b/a$ axis ratio by finding the radius
along the minor axis at which the mean flux is equal to the mean flux at
radius $R_{90}$ along the major axis. The ratio between these radii
is $b/a$. The derived axis ratio is relatively insensitive to the
radius at which it is measured.

In Figure~\ref{cpetro-vs-ba}, we have calculated \cpetro\ and $b/a$ for
various values of $B/D$ from $0$ to $10$. In each case, we use
$R_e = 0.15 R_d$ \citep{balcells-etal07},
disc thickness $z_h = 0.2 R_d$,
and calculate the quantities at inclinations
$i=0\degr, 2.5\degr, 5\degr, \ldots 87.5\degr, 90\degr$.

\section{Subtleties of the Petrosian Concentration Index}
\label{concentration-section}

While modelling the concentration of bulge+disc galaxies, we have
discovered several unintuitive behaviours of the
Petrosian concentration index. We discuss two of these here:
in Section~\ref{concentration-discs}, we investigate
why observed late-type disc galaxies have lower values of \cpetro\ than
idealized exponential discs, while
in Section~\ref{concentration-bulgy}, we investigate why the
concentration of bulge+disc systems is not monotonic with bulge-to-disc
ratio.

\subsection{The low concentration of disc galaxies}
\label{concentration-discs}
In Figures~\ref{cpetro-vs-ba} and \ref{cnorm-vs-ba}, the observed late-type
disc galaxies have lower concentrations than expected for a pure
exponential disc with no bulge (the $B/D=0$ curve and $\cnorm=1$
respectively). In the na\"{i}ve picture in which a galaxy is composed
of a superposition of a low-concentration disc and a high-concentration
bulge, it is impossible to construct a galaxy with $\cnorm<1$.
Why, then, is this the region of parameter space populated by disc
galaxies?

\begin{figure}
\includegraphics[scale=0.5]{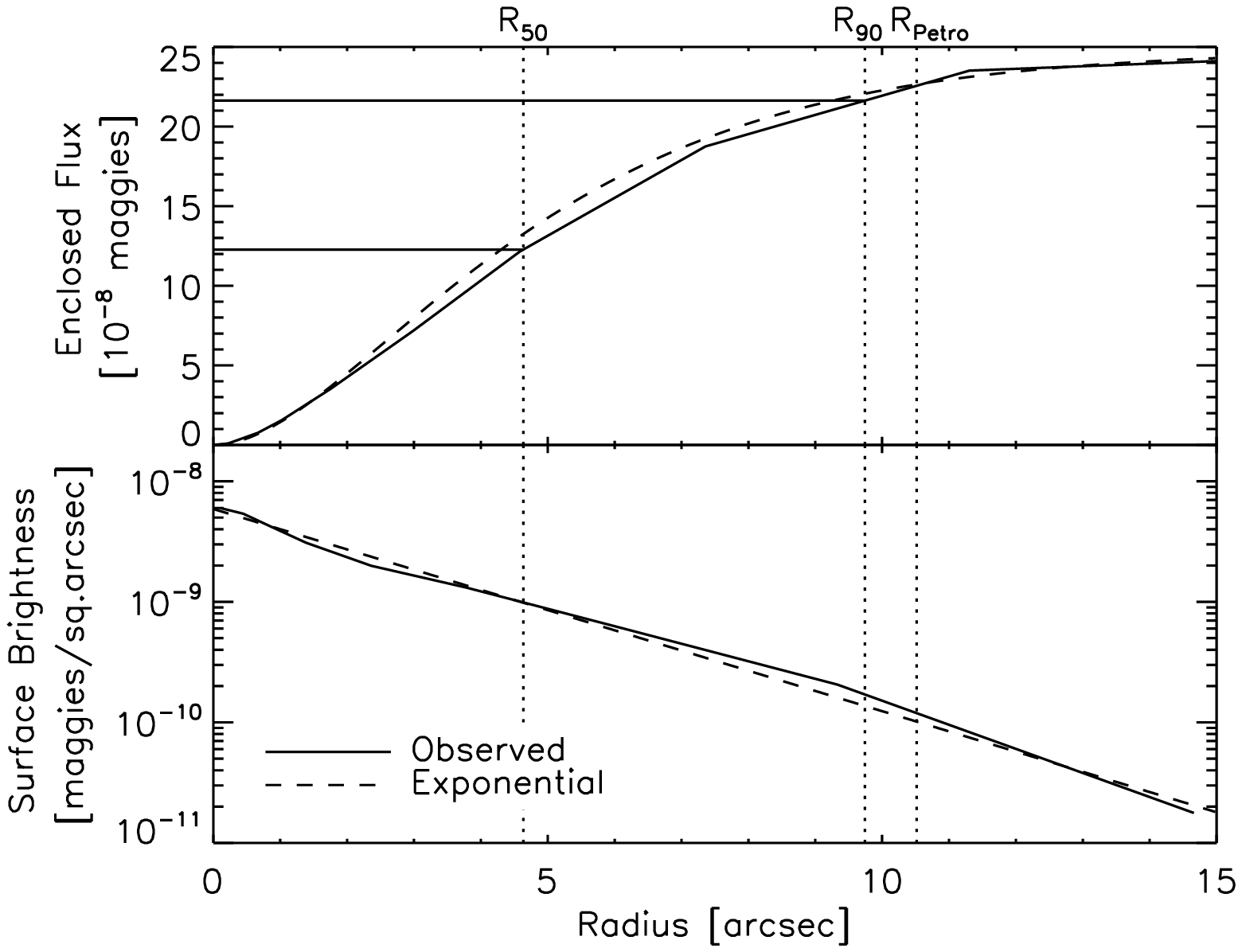}
\caption{\label{sample-profile}%
The r-band surface brightness profile (bottom) and enclosed flux profile (top)
of the sample disc galaxy shown in Figure~\ref{postage-stamps} with
$\cnorm=0.9$ and $b/a=0.7$. The solid curve is the observed
profile, while the dashed curve indicates the best fit pure exponential
profile. The Petrosian radius $R_{\mathrm{Petro}}$ and the $50\%$ and
$90\%$ radii $R_{50}$ and $R_{90}$ are indicated.}
\end{figure}

To investigate this, we have plotted in Figure~\ref{sample-profile}
the surface brightness profile and enclosed flux profile of a sample
galaxy in this region of parameter space (it is the galaxy at
$\cnorm=0.9$, $b/a=0.7$ in Figure~\ref{postage-stamps}), along with
the best-fit exponential profile, which would have $\cnorm=1$.
It can be seen that the observed profile does not follow the pure
exponential, but rather has three features: a central concentration
(at $r<2\arcsec$), an exponential disc that is shallower than the
best-fit exponential ($2\arcsec < r < 10\arcsec$), and a steeper
exponential at large radius. Such outer truncations are well-known
observationally, especially among the latest-type spiral
galaxies \citep{pt06}.

The $90\%$ radius, $R_{90}$, is determined by the global size of
the disc, and is therefore not a strong function of the detailed
flux distribution. This can be seen by how similar the observed (solid)
and idealized exponential (dashed) enclosed flux
curves are in the region surrounding $R_{90}$. However, $R_{50}$ lies
in a region where the observed surface brightness profile is significantly
shallower than the best-fit exponential, and therefore $R_{50}$ is
pushed to larger radii. This combination decreases the measured concentration
relative to that of a pure exponential, resulting in $\cnorm < 1$.
Such profiles are extremely common amongst disc galaxies, and therefore
the late-type disc galaxy locus is centred at $\cnorm=0.9$ rather than
$\cnorm=1$ as one would na\"{i}vely expect.

\subsection{The high concentration of bulge-dominated disc galaxies}
\label{concentration-bulgy}
In Figure~\ref{cpetro-vs-ba}, we have plotted the relation
between \cpetro\ and $b/a$ for idealized bulge+disc systems with
$B/D=0$, $3$ and $10$. The most bulge-dominated of these
shows a curious behaviour: the concentration is significantly higher
than for pure bulge systems, especially when the disc is face-on%
\footnote{The same effect can be seen in figure~5 of
\citet{blanton-etal03-properties}, although they do not comment on it}.
How can the addition of a $10\%$ component that is \textit{more} extended
than the bulge result in a \textit{higher} measured concentration?

\begin{figure}
\includegraphics[scale=0.5]{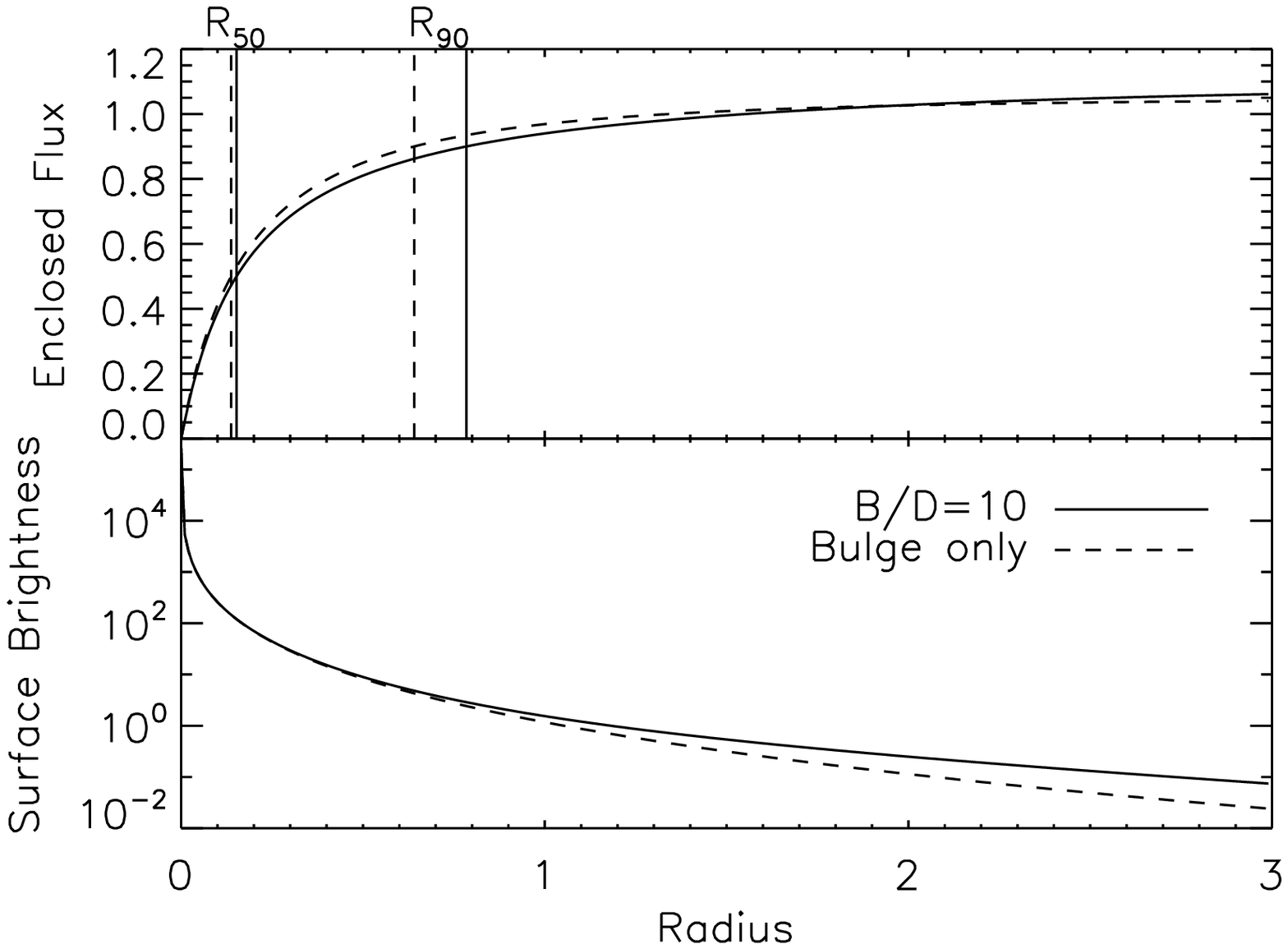}
\caption{\label{bd-profile}%
The surface brightness profile (bottom) and enclosed flux profile (top) of
model bulge-dominated galaxies. The solid curves indicate a galaxy with a
bulge-to-disc ratio, $B/D$, of $10$, while the dashed curves indicate a pure
bulge galaxy. Enclosed flux is given in units of the Petrosian flux, surface
brightness is in arbitrary units, and
radii are in units of the disc scale length $R_d$, where the
bulge effective radius is $R_e = 0.15 R_d$. The $50\%$ and $90\%$ radii,
$R_{50}$ and $R_{90}$ are indicated.}
\end{figure}

To investigate this, we have plotted in Figure~\ref{bd-profile}
the surface brightness and enclosed flux profiles of both a pure
bulge and a face-on bulge+disc system with $B/D=10$. In both cases the
bulge effective radius $R_e=0.15$. For the bulge+disc system,
the disc scale length is $R_d=1$. The surface brightness profiles are
normalized to the same bulge luminosity, while the enclosed flux profiles
are normalized to the Petrosian flux in each case.

Because of the high $B/D$ ratio, the bulge dominates the total luminosity.
Therefore, the $50\%$ and $90\%$
flux levels that define $R_{50}$ and $R_{90}$ are determined by the bulge.
The bulge is also much more concentrated than the disc and completely
dominates the flux inside the effective radius $R_e$, which contains
half the bulge flux and therefore determines $R_{50}$.
However, there is a region of the surface brightness profile that is
dominated by the disc. The flux distribution in this region is
more extended than it would be in the absence of the disc, and therefore
radii enclosing a fixed fraction of the flux
\textit{increase} due to the presence
of the extended component.
This is precisely the situation for $R_{90}$ in Figure~\ref{bd-profile}.
Therefore, the presence of the disc increases $R_{90}$ without
affecting $R_{50}$, resulting in a higher concentration!

The conditions required for this behaviour are
\begin{enumerate}
 \item the bulge dominates the total luminosity,
 \item the bulge dominates the surface brightness profile within $R_{50}$,
and
 \item the disc dominates the surface brightness profile at $R_{90}$.
\end{enumerate}
While these may appear to
be quite restrictive conditions, they are met for very reasonable
sets of parameters.
The effect is less pronounced for inclined discs precisely because an inclined
disc component appears less extended, and therefore has less of an
effect on $R_{90}$.

As a result, \cpetro\ and \cnorm\ are not monotonic functions of $B/D$,
and in particular the galaxies with the largest concentrations are not
pure spheroidal galaxies (although they are bulge-dominated).

\end{document}